\documentclass[12pt,a4paper]{article}

\usepackage{epsfig}
\usepackage{amsmath,amsfonts, amssymb}

\begin{document}
\bibliographystyle{plain}
\pagenumbering{arabic}
\newcommand{\tr}[1]{\mbox{Tr} \, #1 }
\newtheorem{theorem}{Theorem}
\newtheorem{lemma}[theorem]{Lemma}
\newtheorem{prop}[theorem]{Proposition}
\newtheorem{example}[theorem]{Example}
\def \iden {{\bf{I}}}
\def\a{{\lambda_1}}

\renewcommand{\theequation}{\arabic{equation}}
\renewcommand{\thetheorem}{\arabic{theorem}}
\renewcommand{\thesection}{\arabic{section}}

\newenvironment{proof}
{\noindent{\bf Proof\ }}{{\hfill \endbox }\par\vskip2\parsep}

\newcommand{\conpr}{\buildrel{\pr}\over\longrightarrow}
\newcommand{{\Tr}}{\rm Tr}

\newcommand{\RR}{{\mathbb R}}
\newcommand{\pr}{{\mathbb P}}
\newcommand{\ZZ}{{\mathbb Z}}
\newcommand{\NN}{{\mathbb N}}
\newcommand{\CC}{{\mathbb C}}

\newcommand{\hilb}{{\cal{H}}^{\otimes n}}

\newcommand{\EE}{{\mathbb E}}
\newcommand{\ov}[1]{\overline{#1}}

\parindent=0pt \parskip=10pt

\setcounter{equation}{0}

\newcommand \qed {\vrule height5pt width5pt}
\newcommand{\be}{\begin{equation}}
\newcommand{\ee}{\end{equation}}
\newcommand{\bea}{\begin{eqnarray}}
\newcommand{\eea}{\end{eqnarray}}
\newcommand \boundary {\partial}
\newcommand \eps {\epsilon}
\newcommand \half {\frac{1}{2}}
\newcommand \one {{\mbox{\bf 1}}}
\newcommand{\restrict}{{\upharpoonright}}
\newcommand \x {{\underline{x}}}
\def\reff#1{(\ref{#1})}
%

\font\titlefnt=cmr6 scaled \magstep3

\title{\bf{
Additivity in Isotropic Quantum Spin Channels}}

\author{Nilanjana Datta 
\\Statistical Laboratory
\\Centre for Mathematical Sciences
\\University of Cambridge
\\Wilberforce Road, Cambridge CB30WB
\\ email: n.datta@statslab.cam.ac.uk
}

\maketitle

\begin{abstract}
We prove additivity of the minimum output entropy and the Holevo capacity
for rotationally invariant quantum channels acting on 
spin--$1/2$ and spin--$1$ systems. The physical 
significance of these channels and their relations to other known 
channels is also discussed. 
\end{abstract}
\section{Introduction}
\label{introduction}
Information is transmitted through channels which are inherently noisy. 
A measure of the efficiency of a channel is hence given by the maximal rate at 
which information can be reliably (that is, without distortion) transferred 
through it. For a classical communications channel, 
this rate defines its capacity. In contrast, there are various different
capacities of a quantum channel. This is because the structure of
a quantum channel, $\Phi$, is much more complex than its classical 
counterpart, and in addition, there is a lot of flexibility in its use.
A quantum channel can  be used to transmit either classical 
or quantum information; there are various ways of encoding (decoding) the 
input (output) of such a channel; and it can be 
used in conjunction with additional resources e.g. shared entanglement 
between the sender and the receiver. There 
are various quantities characterizing a quantum channel, 
which can be used to describe how efficiently information can be 
transmitted through it. Two such quantities are the minimum output 
entropy $h(\Phi)$ and the Holevo capacity $\chi(\Phi)$ of the channel. 

Mathematically, a quantum channel $\Phi$ is defined as a completely positive 
trace--preserving map on the set of density matrices $\rho$ 
acting on a Hilbert space $\mathcal{H}$. This definition can be extended to the $*$--algebra ${\cal{B}}({\mathcal{H}})$ 
of all complex $d \times d$ matrices, $d < \infty$ being 
the dimension of $\mathcal{H}$. 
A quantum channel $\Phi$ is said to be memoryless if its repeated
use is given by the tensor product 
$\Phi \otimes \Phi \otimes \cdots \otimes \Phi$, since in this case the action 
of each use of the channel is identical, and it is independent for 
different uses.
 
The minimum output entropy of a channel $\Phi$ is defined as
\begin{equation}
h(\Phi ):=\min_{\rho }\,S(\Phi (\rho )),  \label{minent1}
\end{equation}
where the minimization is over all possible input states of the channel.
Here $S(\sigma )= -\mbox{tr}(\sigma \log \sigma)$, the von Neumann entropy 
of the output $\sigma =\Phi (\rho) $ 
 of the channel. 
The von Neumann entropy $S(\sigma )$ is zero if and only if 
$\sigma$ is a pure state density matrix. Hence, the minimum output
entropy gives a measure of the extent to which the output
of a channel deviates from a pure state.

The Holevo capacity of a memoryless channel $\Phi$ is defined as 
\be
\chi(\Phi )=\max_{p_j,\rho_j}\left[S\left(\Phi\left(\sum_j p_j\rho_j\right)
\right) 
- \sum_j p_j S\left(\Phi\left(\rho_j\right)\right)\right],
\label{cap}
\ee
where the maximum is over all finite ensembles of states $\rho_j$ taken with
probabilites $p_j$. It is known to represent the
asymptotic capacity of the quantum channel for transmission of classical information through it, under the restriction that each input state used 
can be expressed as a 
tensor product (see e.g. \cite{hol0}). 

It is conjectured (see e.g. \cite{hol}) that $h(\Phi )$ and $\chi(\Phi )$ are additive for 
general product
channels, i.e., for any two channels $\Phi_1$ and $\Phi_2$
\be
h(\Phi_1 \otimes \Phi_2) = h(\Phi_1) + h(\Phi_2),
\label{addmoe}
\ee
and 
\be
\chi(\Phi_1 \otimes \Phi_2) = \chi(\Phi_1) + \chi(\Phi_2) .
\label{addcap}
\ee
The product channel $\Phi_1 \otimes \Phi_2$, for $\Phi_1=\Phi_2 =\Phi$,
describes two consecutive uses of a memoryless channel $\Phi$. In this
case the additivity relations \reff{addmoe} and \reff{addcap} reduce to
\be
h(\Phi \otimes \Phi) = 2 h(\Phi) \,\,\,\,\quad (a)\,\, ; \quad 
\chi(\Phi \otimes \Phi) = 2 \chi(\Phi). \,\,\,\,\quad(b)
\label{addrel}
\ee
These have a natural generalisation to additivity under 
$n$ uses of the channel:
\be
h(\Phi^{\otimes n}) = n h(\Phi) \quad ; \quad 
\chi(\Phi^{\otimes n}) = n \chi(\Phi).
\label{addreln}
\ee

In this paper we prove the additivity relations \reff{addrel} 
for isotropic spin channels. These were  
first introduced in the context of Quantum Information Theory in 
\cite{alicki}, and are given by the following
bistochastic completely positive maps acting on spin--$s$ systems:
\be
\Phi_s(\rho) = \frac{1}{s(s+1)} \sum_{k=1}^3 S_k \rho S_k.
\label{ch1}
\ee
Here $S_1,S_2,S_3$ are spin operators which satisfy the
commutation relations $[S_i, S_j] = i \epsilon_{ijk} S_k$, 
and provide an irreducible representation 
$\{{U_s({\bf{n}}), {\bf{n}} \in {\mathbb{R}}^3, |{\bf{n}}| \le 1}\}$ of 
the SU(2) group on a Hilbert space of dimension $d= 2s + 1$.
Moreover, they belong to a family of channels referred to as Casimir
channels \cite{greg}, which can be defined starting from any compact 
Lie group. Here we focus attention on the cases $s=1/2$ and $s=1$ 
of the channel $\Phi_s$.

The channel $\Phi_s$ satisfies the following covariance (or rotational invariance)
relation
\be
\Phi_s\bigl(U_s({\bf{n}}) \rho U^*_s({\bf{n}})\bigr) = U_s({\bf{n}})\Phi_s(\rho) 
U^*_s({\bf{n}}).
\label{cov}
\ee
This property leads to certain simplifications in proving additivity relations
for these channels. This is because for channels satisfying \reff{cov}, the
minimum output entropy $h(\Phi_s)$ and the Holevo capacity $\chi(\Phi_s)$ are
linearly related \cite{hol1}:
$$
\chi(\Phi_s) = \log d - h(\Phi_s).
$$
Therefore, the additivity (\ref{addrel}a) of $h(\Phi_s)$ implies the additivity 
(\ref{addrel}b) for such channels.

For $s=1/2$, the above channel reduces to a unital qubit channel:
\be
\Phi(\rho):= \Phi_{1/2}(\rho)  = \frac{1}{3} \sum_{k=1}^3 \sigma_k \rho \sigma_k,
\label{ch2}
\ee
where $\sigma_1, \sigma_2, \sigma_3$ are the three Pauli matrices
$\sigma_x$, $\sigma_y$ and $\sigma_z$ respectively.
The additivity of the minimum output entropy and the Holevo capacity for unital qubit channels was proved in \cite{king}. Our paper gives an 
alternative proof of the additivity relations \reff{addrel} 
for the product channel $\Phi \otimes \Phi$, based on 
the method developed in \cite{dhs}. It provides an explicit illustration
of the sufficient condition for this additivity, which was 
given in \cite{suff} (see Section \ref{method} for more details).

The channel \reff{ch2} can be viewed as the familiar depolarizing channel, 
in the limit in which a qubit is subject to errors represented by  
the Pauli matrices $\sigma_x$, $\sigma_y$ and $\sigma_z$, each with probability
$1/3$. The channel has yet another significance. It is known that an 
unknown quantum state cannot be perfectly copied (no--cloning theorem, \cite{buzek1})
or perfectly complemented \cite{buzek2}.  In \cite{buzek2}, Buzek et al. defined
the Universal--NOT (U--NOT) gate, which generates $M$ output qubits in a state which is as close as possible to the perfect
complement of the states of $N$ identically prepared input qubits.
The channel \reff{ch2} can be shown to provide the optimal realization 
of the U--NOT gate for $N=1$. This can be seen as follows. The state of a 
qubit is given in the Bloch representation as 
$\rho = \frac{1}{2} \left( {\iden} + {\vec{s}} . {\vec{\sigma}}\right)$,
where $\vec{s} = \left( s_1, s_2, s_3\right)$ is its Bloch vector. The angle between the 
Bloch vectors of the state $\rho$ and its complement $\rho^\perp$ is $\pi$. In fact, if 
$\rho$ is a pure state, the points corresponding to $\rho$ and $\rho^\perp$ are
antipodes on the Bloch sphere. It follows from \cite{buzek2} that for $N=1$, the 
optimal U--NOT gate converts the input state $\rho$ of a qubit to the state
\be
\rho^\perp + \frac{1}{3} {{\iden}} \equiv \frac{1}{2} \left( {{\iden}} + {\vec{s'}}. {\vec{\sigma}}\right),
\label{outunot} 
\ee 
where ${\vec{s'}} = - (1/3) \vec{s}$. In other words the gate causes a shrinking 
of the Bloch vector (by a factor $1/3$) in addition to a rotation by an angle $\pi$.
Using the Bloch representation of $\rho$ we find that the output $\Phi(\rho)$ 
of the channel \reff{ch2} is indeed given by \reff{outunot}. 

The channel \reff{ch1} for $s=1$ is given by 
\be
 \Phi_1 (\rho) = \frac{1}{2} \sum_{k=1}^3 S_k \rho S_k,
\label{ch3}
\ee
where $S_k$, $k=1,2,3$ are the corresponding spin operators. The underlying
Hilbert space is three--dimensional. 
We show in Section \ref{last} that $\Phi_1$ is equivalent to the
channel ${\widetilde{\Phi}}_d$ defined by 
\be
{\widetilde{\Phi}}_d(\mu) := \frac{1}{d-1} \bigl({\iden}\,\tr \mu - \mu^T \bigr),
\label{ours}
\ee   
{\em{for the choice}} $d=3$. Here $\mu \in {\cal{B}}({\mathcal{H}})$,
where ${\cal{H}} \simeq {\mathbf{C}}^{d}$, 
$\mu^T$ denotes the transpose of the matrix $\mu$, and ${\iden}$ denotes
the $d \times d$ unit matrix. 
The channels $\Phi_1$ and ${\widetilde{\Phi}}_3$ are equivalent 
in the sense that, for any $\mu \in {\mathbf{C}}^{3}$, $\Phi_1(\mu)=
{\widetilde{\Phi}}_3(\mu)$.
The additivity relations \reff{addreln} 
for the minimum output entropy and the Holevo capacity of the 
channel ${\widetilde{\Phi}}_d$, for any arbitrary dimension 
$2 \le d < \infty$, was proved in 
\cite{my, af}. Additivity \reff{addrel} for the product channel
was also proved in \cite{dhs}. 
The equivalence between the channels $\Phi_1$ and 
${\widetilde{\Phi}}_3$ therefore imply the validity of these additivity 
relations for $\Phi_1$ as well.
 
The paper is organized as follows. In Section \ref{method}
 we briefly summarize
the key idea behind the method developed in \cite{dhs}, with 
emphasis on the sufficient condition for additivity of the minimum output
entropy of product channels. In Section \ref{unot} we give 
an explicit proof of the additivity relation \reff{addrel} for the channel
\reff{ch2}, i.e., for the case $s=1/2$. In Section \ref{last} 
the equivalence between the 
channels $\Phi_1$ and ${\widetilde{\Phi}}_3$ is proved by two different
methods. The additivity relations \reff{addreln} follow as a consequence.  


\section{A sketch of the method}
\label{method}
Our aim is to prove the additivity relations \reff{addrel} for product
channels $\Phi \otimes \Phi$, for the cases in which $\Phi$ is defined through
\reff{ch1} and \reff{ch2}. The concavity of the von Neumann
entropy implies that the minimum value of the output entropy is necessarily 
achieved for pure input states, which correspond to the extreme 
points of the convex set of input states. This allows the 
minimisations in the definitions \reff{minent1} of the minimum output 
entropy to be restricted to pure input states
alone. Hence, the minimum output entropies of a single channel $\Phi$
on ${\cal{B}}({\mathcal{H}})$, and the 
product channel $\Phi \otimes \Phi$ on ${\cal{B}}(\mathcal{H} 
\otimes {\mathcal{H}})$, is equivalently given by
\begin{eqnarray}
h(\Phi ) &=&\min_{{{{|\psi\rangle \in \mathcal{H}}}}{{, ||\psi ||=1} }
}\,S(\Phi (|\psi\rangle\langle \psi |)),\quad  \label{entrop1} \\
h(\Phi\otimes \Phi) &=& \min_{{{{|\psi_{12}\rangle \in {
\mathcal{H}}_1\otimes {\mathcal{H}}_2 }}}{{{,\ ||\psi_{12} ||=1}}}}\,
S((\Phi\otimes \Phi)(|\psi_{12}\rangle\langle \psi_{12}|)).
\label{entrop2}
\end{eqnarray}
Here $|\psi_{12}\rangle\langle \psi_{12}|$ is a pure state of a bipartite
system with the Hilbert space ${\mathcal{H}}_1\otimes {\mathcal{H }}_2$,
where ${\mathcal{H}}_1\simeq {\mathcal{H }}_2 \simeq {\mathcal{H}}$.
In order to prove (\ref{addrel}a), it is sufficient to show that the
minimum in \reff{entrop2} is achieved on unentangled vectors
$|\psi _{12}\rangle =|\psi _{1}\rangle \otimes |\psi _{2}\rangle \in {
\mathcal{H}}_1\otimes {\mathcal{H}}_2$. This follows from the fact that the 
von Neumann entropy of a tensor product state is additive.

The starting point of our analysis is the Schmidt decomposition
\begin{equation}
|\psi _{12}\rangle =\sum_{\alpha =1}^{d}\sqrt{\lambda _{\alpha }}|\alpha
;1\rangle |\alpha ;2\rangle, 
  \label{schmidt}
\end{equation}
where $\{|\alpha ;i\rangle\}$ denotes an orthonormal basis 
in the Hilbert space ${\mathcal{H}}_i$, $i=1,2$. 
The Schmidt coefficients 
form a probability distribution:
$$ \lambda _{\alpha }\geq 0\quad ;\quad \sum_{\alpha =1}^{d}\lambda _{\alpha
}=1,$$
and hence the Schmidt vector 
${\underline{\lambda }} :=(\lambda_1, \ldots, \lambda_d)$ varies in the 
$({d}-1)-$dimensional
simplex $\Sigma _{d}$, defined by these constraints.
 
In terms of the Schmidt decomposition, the input to the product channel is given by 
\begin{equation}\label{schmidt2}
|\psi _{12}\rangle \langle \psi _{12}|=\sum_{\alpha ,\beta =1}^{d}\sqrt{
\lambda _{\alpha }\lambda _{\beta }}|\alpha ;1\rangle \langle \beta
;1|\otimes |\alpha ;2\rangle \langle \beta ;2|,
\end{equation}
and its output is given by 
\begin{equation}
\sigma_{12}({\underline{\lambda }}):=\left( \Phi \otimes \Phi \right)
\left( |\psi _{12}\rangle \langle \psi _{12}|\right) =\sum_{\alpha ,\beta
=1}^{d}\sqrt{\lambda _{\alpha }\lambda _{\beta }}\Phi (|\alpha ;1\rangle
\langle \beta ;1|)\otimes \Phi (|\alpha ;2\rangle \langle \beta ;2|).
\label{matrix}
\end{equation}

Note that the extreme points
(vertices) of $\Sigma _{d}$ correspond to unentangled vectors $
|\psi _{12}\rangle =|\psi _{1}\rangle \otimes |\psi _{2}\rangle \in {
\mathcal{H}}\otimes {\mathcal{H}}$. Hence, a {\em{sufficient condition}}
for the additivity of the minimum output entropy can be stated as 
follows:
\medskip

\noindent
\bea
&& {\hbox{\em{For every choice of the bases }}} \left\{ |\alpha ;1\rangle \right\}  and \left\{ |\alpha ;2\rangle \right\}
 {\hbox{\em{ in }}} 
\reff{matrix} 
, {\hbox{\em{ the von Neumann entropy, }}} 
\nonumber\\
&&
S\left( \sigma_{12}({\ \underline{
\lambda }})\right),  {\hbox{\em{ of the channel output }}} \sigma_{12}({\ \underline{
\lambda }}), {\hbox{\em{ attains its minimum at the vertices of }}}
\Sigma_d.   \nonumber\\
\label{suffcond}\eea 
\medskip

We show in Section \ref{unot}, 
that this sufficient condition for additivity indeed holds
for the channel $\Phi_{1/2}(\rho)$ 
(defined
 by \reff{ch2}). It
is a consequence of the fact that in this case $S\left( \sigma_{12}({\ \underline{
\lambda }})\right)$ is a concave function of the Schmidt vector 
${\underline{\lambda }}$.
For the channel $\Phi_{1}(\rho)$ (defined by \reff{ch3}) as well  
the sufficient condition \reff{suffcond} is satisfied and the additivity 
relations \reff{addrel} hold. This is proved in Section \ref{last} 
by showing that $\Phi_{1}$ is equivalent to a channel for which 
the additivity relations have been previously established.
\section{Additivity for the channel $\Phi_{1/2}$}
\label{unot}

In this section we explicitly prove the validity of the condition
\reff{suffcond}, and hence of the additivity relations \reff{addrel}, 
for the channel $\Phi_{1/2}$ defined by \reff{ch2}. 

Let us first compute the von Neumann entropy 
$S\left[\Phi\right] \equiv S\left(\Phi(\rho)\right)$ of the output
of a single channel $\Phi := \Phi_{1/2}$ :
\bea
\Phi(\rho) &=& \frac{1}{3} \sum_{k=1}^3 \sigma_k \rho \sigma_k =
\frac{1}{3} \sum_{k=1}^3 \sigma_k \left(\frac{1}{2} \sum_{i=0}^3
s_i \sigma_i\right) \sigma_k \nonumber\\
&=& \frac{1}{2}{{\iden}} - \frac{1}{6} \sum_{i=1}^3 s_i \sigma_i,
\label{long1}
\eea
The last line follows from the fact that for $i=1,2,3$, 
\be
\sum_{k=1}^3 \sigma_k \sigma_i \sigma_k = - \sigma_i.
\label{rel1}
\ee
The above identity \reff{rel1} 
can be easily obtained by explicit computation. Hence, $\Phi(\rho)$
is given by the matrix
\begin{equation}
\left(
\begin{array}{cc}
{1}/{2} - {s_3}/{6} & -{s_1}/{6} + i {s_2}/{6}  
\\
-{s_1}/{6} - i {s_2}/{6} & {1}/{2} + {s_3}/{6}\\
\end{array}
\right) .  \label{matrixa}
\end{equation}
To compute $S\left[\Phi\right]$ we need to compute the eigenvalues of
the above matrix. The characteristic equation is given by
\be
\lambda^2 - \lambda + \frac{1}{4} - \frac{1}{36} \left(
s_1^2 + s_2^2 + s_3^2\right) = 0
\ee
which reduces to
\be
\lambda^2 - \lambda + \frac{2}{9} = 0
\ee
since $s_1^2 + s_2^2 + s_3^2 =1 $.
The eigenvalues are found to be equal to $2/3$ and $1/3$. This yields
\be
S\left[\Phi\right] = \log 3 - \frac{2}{3} \log 2
\label{sphi}.
\ee

Let us now calculate the entropy of the output $\sigma_{12} 
\left({\underline{
\lambda }}\right)$ of the product channel $\Phi_{1/2} \otimes \Phi_{1/2}$ .
Note that for the channel $\Phi_{1/2}$ (i.e., for $s=1/2$),
the covariance relation \reff{cov} reduces to 
invariance with respect to any 
arbitrary unitary transformation $U$:
\begin{equation}
\Phi (U\rho U^{\ast })={U}\Phi (\rho ){U^*}.  \label{inv}
\end{equation}
This allows us to choose for 
$\left\{ |\alpha ; j\rangle \,;\, j=1,2 \right\}$
in \reff{schmidt}, the canonical basis in ${\mathbf{C}}^{2}$. Hence, we write
$$
| \psi _{12}\rangle =\sum_{\alpha =1}^{2}\sqrt{\lambda _{\alpha }}|\alpha
\rangle |\alpha \rangle,  \quad {\hbox{ ; }}
|\psi _{12}\rangle \langle  \psi _{12}|=\sum_{\alpha ,\beta =1}^{2}\sqrt{
\lambda _{\alpha }\lambda _{\beta }}|\alpha\rangle \langle \beta 
|\otimes |\alpha \rangle \langle \beta|.
$$
and
\be
\sigma_{12}(\lambda) 
= \sum_{\alpha ,\beta =1}^{2}\sqrt{\lambda _{\alpha
}\lambda _{\beta }}\Phi(|\alpha \rangle \langle \beta  |)\otimes
\Phi (|\alpha \rangle \langle \beta |).  \label{matrixb}
\end{equation}

Let us first compute $\Phi(|\alpha \rangle \langle \beta  |)$.
For this we express the basis vectors $|\alpha \rangle$ in terms
of the vectors $|0\rangle$ and $|1\rangle$ of the computational
basis:
$$
|\alpha \rangle = \sum_{i=0}^1 U_{\alpha i} |i\rangle \quad ; \quad
\langle \beta  | = \sum_{j=0}^1\langle j | {\overline{U}}_{\beta j},
$$
where $U_{\gamma i}$ denote the elements of a unitary matrix and
${\overline{U}}_{\gamma i}$ denotes their complex conjugates.
Hence
\bea
\Phi(|\alpha \rangle \langle \beta |)
&=& \frac{1}{3} \sum_{k=1}^3 \sigma_k |\alpha \rangle \langle 
\beta  | \sigma_k = 
 \frac{1}{3} \sum_{k=1}^3 \sum_{i,j=0}^1\sigma_k 
\left\{ U_{\alpha i} |i\rangle\langle j | {\overline{U}}_{\beta j}\right\} \sigma_k
=  \frac{1}{3}\sum_{i,j=0}^1 U_{\alpha i} {\overline{U}}_{\beta j} \left(
\sum_{k=1}^3\sigma_k |i\rangle\langle j |\sigma_k\right) \nonumber\\
&=&  \frac{1}{3}\Bigl[ U_{\alpha 0} {\overline{U}}_{\beta 0} \bigl(
\sum_{k=1}^3\sigma_k |0\rangle\langle 0|\sigma_k\bigr)
 + U_{\alpha 1} {\overline{U}}_{\beta 1} \bigl(
\sum_{k=1}^3\sigma_k |1\rangle\langle 1|\sigma_k\bigr)
\nonumber\\
 & & 
+ U_{\alpha 0} {\overline{U}}_{\beta 1} \bigl(
\sum_{k=1}^3\sigma_k |0\rangle\langle 1|\sigma_k\bigr)
+ U_{\alpha 1} {\overline{U}}_{\beta 0} \bigl(
\sum_{k=1}^3\sigma_k |1\rangle\langle 0|\sigma_k\bigr)\Bigr].\nonumber\\
\label{main1}
\eea
Now,
\bea
\sum_{k=1}^3\sigma_k |0\rangle\langle 0|\sigma_k
= 2{\iden} - |0\rangle\langle 0|  \,&;&\, \sum_{k=1}^3\sigma_k |1\rangle\langle 1|\sigma_k
= 2{\iden} - |1\rangle\langle 1|
\nonumber\\
\,&;&\,
\sum_{k=1}^3\sigma_k |0\rangle\langle 1|\sigma_k
= - |0\rangle\langle 1| \,;\,
\sum_{k=1}^3\sigma_k |1\rangle\langle 0|\sigma_k
= - |1\rangle\langle 0|,
\nonumber\\
\eea
Substituting the above relations in the RHS of \reff{main1} yields
\bea
\Phi(|\alpha \rangle \langle \beta  |)
&=& \frac{1}{3} \Bigl[2{\iden}(U_{\alpha 0} {\overline{U}}_{\beta 0} +
U_{\alpha 1} {\overline{U_{\beta 1}}}) - U_{\alpha 0} {\overline{U}}_{\beta 0} |0\rangle\langle 0|
- U_{\alpha 1} {\overline{U}}_{\beta 1} |1\rangle\langle 1| \nonumber\\
& & \quad - U_{\alpha 0} {\overline{U}}_{\beta 1} |0\rangle\langle 1|
- U_{\alpha 1} {\overline{U}}_{\beta 0} |1\rangle\langle 0| \Bigr]\nonumber\\
&=& \frac{1}{3} \Bigl[2{\iden} \delta_{\alpha\beta} -
|\alpha\rangle\langle \beta|\Bigr].
\label{main2}
\eea
The last line follows from the relation
\be
|\alpha\rangle\langle \beta| = U_{\alpha 0} {\overline{U}}_{\beta 0}|0\rangle\langle 0|
+ U_{\alpha 1} {\overline{U}}_{\beta 1} |1\rangle\langle 1| + 
U_{\alpha 0} {\overline{U}}_{\beta 1} |0\rangle\langle 1|
+ U_{\alpha 1} {\overline{U}}_{\beta 0} |1\rangle\langle 0|. 
\label{main3}
\ee
Substituting \reff{main2} on the RHS of \reff{matrixb} we get
\bea
\sigma_{12}({\underline{\lambda }}) &=& \frac{1}{9} 
\sum_{\alpha ,\beta =1}^{2}\sqrt{\lambda _{\alpha
}\lambda _{\beta }}
\left(2{\iden} \delta_{\alpha\beta} -
|\alpha\rangle\langle \beta|\right) \otimes 
\left(2{\iden} \delta_{\alpha\beta} -
|\alpha\rangle\langle \beta|\right) \nonumber\\
&=& \frac{1}{9} \Bigl(\sum_{\alpha ,\beta=1}^{2}
|\alpha \beta \rangle \langle \alpha \beta |(4-2\lambda _{\alpha
}-2\lambda _{\beta })+\sum_{\alpha ,\beta =1}^{2}\sqrt{\lambda _{\alpha
}\lambda _{\beta }}|\alpha \alpha \rangle \langle \beta \beta |\Bigr) .
\label{opschmidt}
\eea
In the above we have used the completeness relations
\[
\,{\iden}=\sum_{\alpha =1}^{2}|\alpha \rangle \langle \alpha |,\quad {\iden}
\otimes {\iden}=\sum_{\alpha ,\beta =1}^{2}|\alpha \beta \rangle \langle
\alpha \beta |.
\]
The matrix $9 \sigma_{12}({\underline{\lambda }})$can be written in the form of a more general matrix
$$
A=\sum_{j=1}^{n}\mu _{j}|j\rangle \langle j|+\sum_{j,k=1}^{n}\sqrt{\eta
_{j}\eta _{k}}|j\rangle \langle k|,  \label{opa}
$$
by identifying $j$ with a pair $(\alpha ,\beta )$
and setting
\begin{equation}
\mu _{j}\equiv \mu _{\alpha \beta }=4-2\lambda _{\alpha }-2\lambda _{\beta
}\quad ;\quad \eta _{j}\equiv \eta _{\alpha \beta }=\lambda _{\alpha }\delta
_{\alpha \beta },\quad \alpha ,\beta =1,2.  \label{parameters}
\end{equation}
The eigenvalues of the matrix $9 \sigma_{12}({\underline{\lambda }})$ are the roots
of the equation
\bea
0 &=&\prod_{\alpha ,\beta =1}^{2}(4-2\lambda _{\alpha }-2\lambda
_{\beta }-\gamma )\left[1+\sum_{\alpha ^{\prime },\beta ^{\prime }=1}^{2}
\frac{\lambda _{\alpha ^{\prime }}\delta _{{\alpha ^{\prime }}{\beta
^{\prime }}}}{(4-2\lambda _{\alpha^{\prime } }-2\lambda
_{\beta^{\prime } }-\gamma)}
\right]  \nonumber  \label{long} \\
&=&\prod_{{{\alpha ,\beta =1}}{{\alpha \neq \beta }}}^{2}
(4-2\lambda _{\alpha }-2\lambda
_{\beta }-\gamma)\left[\prod_{\alpha ^{\prime
}=1}^{2}(4-4\lambda _{\alpha ^{\prime } }-\gamma))\left\{1+\sum_{\alpha
^{^{\prime \prime }}=1}^{2}\frac{\lambda _{\alpha ^{^{\prime \prime }}}}{
(4-4\lambda _{\alpha ^{^{\prime \prime }}}-\gamma )}\right\}\right].  \nonumber
\\
&&
\label{sec}
\end{eqnarray}

Eq.(\ref{sec}) yields the following equations:
\begin{equation}
(4-2\lambda _{\alpha }-2\lambda _{\beta }-\gamma ) 
\equiv 2 - \gamma =0,\quad \alpha \neq \beta
,\quad \alpha ,\beta = 1,2,  \label{one1}
\end{equation}
since the Schmidt coefficients satisfy 
$\lambda _{\alpha }+ \lambda _{\beta } = 1$. 
Hence the matrix $9 \sigma_{12}({\underline{\lambda }})$ has $2$
eigenvalues equal to $2$ and the matrix $\sigma_{12}({\underline{\lambda }})$
has two eigenvalues equal to $2/9$.

The roots of the equation
\begin{equation}
\prod_{\alpha =1}^{2}(4-4\lambda _{\alpha }-\gamma )\left\{1+\sum_{\alpha'
=1}^{d}\frac{\lambda _{\alpha' }}{(4-4\lambda _{\alpha' }-\gamma )}\right\}=0,
\label{two}
\end{equation}
give the remaining $2$ eigenvalues of the matrix
$9 \sigma_{12}({\underline{\lambda }})$. These are obtained as follows.
Eq.\reff{two} can be written as 
\be
(4-4\lambda _{1}-\gamma )(4-4\lambda _{2}-\gamma )
+ \lambda_1 (4-4\lambda _{2}-\gamma ) + \lambda_2 (4-4\lambda _{1}-\gamma )
= 0,
\ee
which reduces to the quadratic equation
$
\gamma^2 - 5 \gamma + (8\lambda _{1}\lambda _{2} + 4) = 0. 
$
This has roots 
$
\bigl[{5}/{2} \pm \left(\sqrt{ 9 -32 \lambda_1 \lambda_2}\right)/{2}\bigr].$
Hence the corresponding eigenvalues
of the matrix $\sigma_{12}({\underline{\lambda }})$ are 
$
\bigl[{5}/{18} \pm \left(\sqrt{ 9 -32 \lambda_1 \lambda_2}\right)/{18}\bigr]$.

The $4$ eigenvalues of $\sigma_{12}({\underline{\lambda }})$
are therefore given by 
\be
r_1= \frac{2}{9},\,\, r_2 = \frac{2}{9}, \,\,
r_3 = \frac{5}{18} + \frac{\sqrt{ 9 -32 \lambda_1 \lambda_2}}{18}, 
\,\,r_4= \frac{5}{18} - \frac{\sqrt{ 9 -32 \lambda_1 \lambda_2}}{18}. 
\label{eigen}
\ee
The output entropy
$S\bigl(\sigma_{12}({\underline{\lambda }})\bigr) := S 
\left( \Phi _{1}\otimes \Phi _{2}\right) \left(
|\psi _{12}\rangle \langle \psi _{12}|\right)$
of the product channel is given by 
$$
S\bigl(\sigma_{12}({\underline{\lambda }})\bigr) = - \sum_{i=1}^4 r_i \log r_i.
$$

Since $\lambda_2 = 1 - \lambda_1$ we write
$$
f_1(\a) \equiv r_3 := \frac{5}{18} +
\frac{\sqrt{ 9 -32 \lambda_1 (1 - \a)}}{18},$$
and
 $$
f_2(\a) \equiv r_4 := \frac{5}{18} - 
\frac{\sqrt{ 9 -32 \lambda_1 (1 - \a)}}{18}.$$
Hence,
\be 
S\bigl(\sigma_{12}({\underline{\lambda }})\bigr)\equiv 
S\bigl(\sigma_{12}({{\lambda_1 }})\bigr) :=  T_1 + T_2 (\a),
\ee
where $T_1$ is a constant:
\be
T_1 = - (r_1 \log r_1 + r_2 \log r_2) = 
-\frac{4}{9} \log \left(\frac{2}{9}\right)
\ee
and  
\be
T_2(\a) =  - \bigl(r_3 \log r_3 + r_4 \log r_4 \bigr) =
 - \bigl( f_1(\a)\log f_1(\a) + f_2(\a) \log f_2(\a) \bigr). 
\label{concave1}
\ee
Note that $f_1(\a) + f_2(\a) = 10/18$ and ${f_2}^\prime(\a)= -{f_1}^{\prime}(\a)$,
where the prime denotes differentiation with respect to $\lambda_1$.
The sufficient condition \reff{suffcond} is satisfied, if the 
output entropy $S\bigl(\sigma_{12}({\underline{\lambda }})\bigr)$ 
is a concave function of the Schmidt vector $\underline{\lambda }$. 
To prove concavity, it suffices to show that
$$S^{''} \bigl(\sigma_{12}({{\lambda_1 }})\bigr) :=
\frac{d^2 S\bigl(\sigma_{12}({{\lambda_1 }})\bigr)}{d\lambda_1^2} < 0.$$

We find that 
\be
S^{''}\bigl(\sigma_{12}({\underline{\lambda }})\bigr) = T_2^{''}(\a) =
f_1^{''}(\a) \log \bigl[\frac{f_2(\a)}{f_1(\a)}\bigr] - \frac{5}{9}\left(
{f_1}^\prime(\a)\right)^2 \frac{1}{f_1(\a)f_2(\a)}.
\label{der}
\ee
Now,
\be
 f_1^{''}(\a) = 
\frac{16}{9} \left( 9 -32 \lambda_1 (1- \lambda_1)\right)^{-3/2} >0,
\ee
whereas 
${\displaystyle{\log \bigl[\frac{f_2(\a)}{f_1(\a)}\bigr] < 0}},$
since $f_2(\a) < {f_1(\a)}$. Hence, the first term on the RHS of \reff{der}
is negative. Moreover,  
$$f_1(\a)f_2(\a) = \left(\frac{5}{18}\right)^2 - 
\frac{1}{(18)^2}\bigl(9 -32 \lambda_1 (1- \lambda_1) \bigr)
= \frac{1}{(18)^2}\bigl(16 + 32 \lambda_1 (1- \lambda_1) \bigr) >0.$$ 
Therefore
$S^{''}\bigl(\sigma_{12}({\underline{\lambda }})\bigr) <0 $ and the output entropy
is a concave function of the Schmidt coefficients.
The concavity of 
$S\bigl(\sigma_{12}({\underline{\lambda }})\bigr)$ implies that it attains  
its minimal value at the vertices of the simplex defined by 
$$\lambda_1, \lambda_2 > 0 \quad ; \quad 
\lambda_1 + \lambda_2 = 1.$$
Hence the sufficient condition for additivity of the minimum output 
entropy is satisfied and the additivity relation (\ref{addrel}a) therefore
holds.

This can also be explicitly verified as follows. 
At any vertex of this simplex, we have $\lambda_1\lambda_2=0$, and hence the eigenvalues 
in \reff{eigen} reduce to 
$$\frac{2}{9}, \,\, \frac{2}{9}, \,\,\frac{4}{9}, \,\,\frac{1}{9}
$$
and the output entropy is 
\bea
S\bigl(\sigma_{12}({\underline{\lambda }})\bigr)
&=&- \Bigl[2 \frac{2}{9} \log  \frac{2}{9} + \frac{4}{9} \log  \frac{4}{9}
+ \frac{1}{9} \log  \frac{1}{9}\Bigr]\nonumber\\
&=& 2 \log 3 - \frac{4}{3} \log 2 \, \equiv \, 2 \times S\left[\Phi\right],
\eea
where $S\left[\Phi\right]$ is given by \reff{sphi}. This proves the additivity relation (\ref{addrel}a). 

As discussed in the Section \ref{method}, the covariance relation \reff{inv}
implies that the Holevo capacity $\chi(\Phi)$ is also additive for  this
channel, i.e. (\ref{addrel}b) holds.

\section{Additivity for the channel $\Phi_1(\rho)$}
\label{last}
In this section we study the channel 
\be
\Phi_1(\rho) = \frac{1}{2} \sum_{k=1}^3 S_k \rho S_k.
\ee
In the basis $\{|s,m\rangle ; s=1, m= -1, 0, 1 \}$, the spin operators
$S_k$ are represented by the matrices
\begin{equation}
S_1=\frac{1}{\sqrt{2}}\left(
\begin{array}{ccc}
0 & 1 & 0
\\
1 & 0 & 1\\
0 & 1 & 0
\end{array} 
\right) \, ; \quad 
S_2= \frac{1}{\sqrt{2}}\left(
\begin{array}{ccc}
0 & -i & 0
\\
i & 0 & -i\\
0 & i & 0
\end{array} 
\right)
 \, ; \quad 
S_3 = \left(
\begin{array}{ccc}
1 & 0 & 0
\\
0 & 0 & 0\\
0 & 0 & -1
\end{array} 
\right).  
\end{equation}
However, if instead of the basis $\{|s,m\rangle\}$, we use the 
basis $\{|a\rangle, |b\rangle, |c\rangle\}$, where
\be
|a\rangle = - \frac{1}{\sqrt{2}}\left(|1\rangle - |-1\rangle \right)
\,;\, 
|b\rangle = \frac{i}{\sqrt{2}}\left(|1\rangle + |-1\rangle \right)
\,;\, 
|c\rangle = |0\rangle,\ee
the spin operators are represented by the following matrices:
\begin{equation}
S'_1=\left(
\begin{array}{ccc}
0 & 0 & 0
\\
0 & 0 & -i\\
0 & i & 0
\end{array} 
\right) \, ; \quad 
S'_2= \left(
\begin{array}{ccc}
0 & 0 & i
\\
0 & 0 & 0\\
-i & 0 & 0
\end{array} 
\right)
 \, ; \quad 
S'_3 = \left(
\begin{array}{ccc}
0 & -i  & 0
\\
i & 0 & 0\\
0 & 0 & 0
\end{array} 
\right), 
\label{rep}
\end{equation}
which are just the matrices representing the infinitesimal generators
of rotations on the components of a vector. It will prove to be 
convenient for our analysis to use the representation \reff{rep} of the
spin operators. 

Note that the matrices $S'_k$ are related to the matrices $S_1,S_2$ and
$S_3$ via a unitary transformation 
$${\vec{S'}} = V {\vec{S}}V^*, \quad {\hbox{where  }}{\vec{S'}}= (S'_1,S'_2, S'_3)\,;\,
{\vec{S}}= (S_1,S_2, S_3),$$
where
\begin{equation}
V=\left(
\begin{array}{ccc}
- \frac{1}{\sqrt{2}} & 0 & \frac{1}{\sqrt{2}}
\\
- \frac{i}{\sqrt{2}} & 0 & -\frac{i}{\sqrt{2}}\\
0 & 1 & 0
\end{array} 
\right). 
\ee

The output $\sigma_{12}({\underline{\lambda }})$ of the product channel, in this case, is given by \reff{matrix},
with $\Phi$ replaced by $\Phi_1$. To compute it, let us first evaluate
$\Phi (|\alpha ;i\rangle \langle \beta ;i|)$, $i=1,2$. For notational
simplicity we omit the index $i$ and express the basis vectors in terms 
of the basis $\{|-1\rangle, |0\rangle, |1\rangle\}$ as follows:
$$
|\alpha \rangle = \sum_{i=-1}^1 U_{\alpha i} |i\rangle \quad ; \quad
\langle \beta  | = \sum_{j=-1}^1\langle j | {\overline{U}}_{\beta j},
$$
where once again $U_{\gamma i}$ denote the elements of a unitary matrix. We obtain
\be
\Phi(|\alpha \rangle \langle \beta  |)
= \frac{1}{2}\sum_{i,j=-1}^1 U_{\alpha i } {\overline{U}}_{\beta j } \left(
\sum_{k=1}^3 S'_k |i\rangle\langle j |S'_k\right) \nonumber,
\ee
which is analogous to \reff{main1} of the previous section. Using the 
relations
\be
\begin{array}{ccc}
S'_1 |1\rangle = 0, &\,\,S'_1 |0\rangle = i|-1\rangle, &\,\,
S'_1 |-1\rangle = - i |0\rangle,  \nonumber\\
S'_2 |1\rangle = -i |-1\rangle, &\,\,S'_2 |0\rangle = 0, &\,\,
S'_2 |-1\rangle = i |1\rangle,  \nonumber\\
S'_3 |1\rangle = i |0\rangle, &\,\,S'_3 |0\rangle = - i|1\rangle,&\,\,
S'_3 |-1\rangle = 0,
\end{array}
\ee
we obtain
\be
\Phi(|\alpha \rangle \langle \beta  |)
= \frac{1}{2}\left({\iden}\delta_{\alpha \beta} -
{{|\overline{\alpha}\rangle}}\,{{\langle\overline{\alpha} |}} \right),
\label{same}
\ee
where the entries of the vector $|{\overline{\alpha}}\rangle $ are complex
conjugates of the corresponding entries of vector $|{\alpha}\rangle$.
Note that the RHS of \reff{same} is identical to the corresponding 
expression for the channel ${\widetilde{\Phi}}_d$ \cite{HW, dhs} 
{\em{for the choice}} $d=3$. This channel is defined by its action
on any $\mu \in {\mathbf{C}}^{d}$ as follows:
\be
{\widetilde{\Phi}}_d (\mu):= \frac{1}{d-1} \bigl({\iden}\,\tr \mu - \mu^T \bigr),
\label{ours2}
\ee   
and was studied in detail \cite{dhs}.
In the latter,
$|\overline{\alpha}\rangle\,\langle\overline{\alpha} |$ was replaced by ${{|\alpha\rangle}}{{\langle\alpha |}}$
because the basis $\{|\alpha\rangle\}$ was chosen to be real. Eq.\reff{ours} 
therefore implies that the channel $\Phi_1$ is equivalent to the
channel ${\widetilde{\Phi}}_3$ of \reff{ours2}.

There is an alternative way of demonstrating this equivalence between the
two channels. It is easy to see that 
${\widetilde{\Phi}}_3(\rho)$ for the channel 
defined by \reff{ours2}, can be expressed as 
(see e.g. \cite{HW}):
\be
{\widetilde{\Phi}}_3(\rho) = \frac{1}{4} \sum_{i,j=1}^3 B_{(ij)} 
\rho  B_{(ij)}^*,
\ee
where $ B_{(ij)}= |j\rangle\langle i| - |i\rangle\langle j|$ and
$\{|i\rangle, i=1,2,3\}$ denotes a complete set of orthonormal  
basis vectors of the Hilbert space ${\cal{H}} \simeq {\mathbf{C}}^{d}$.
Since $ B_{(ii)}= 0$ for all $i$, and $B_{(ij)}= - B_{(ji)}$
for $j \ne i$, ${\widetilde{\Phi}}_3(\rho)$ can be expressed as follows:
\begin{equation}
{\widetilde{\Phi}}_3(\rho) = \frac{1}{2} \sum_{i}^3 A_{i} \rho  A_{i}^*,
\end{equation}
where $A_1 := B_{(12)}, A_2 := B_{(23)}$ and $A_3 := i B_{(31)}$.
Note that the operators $A_i,\, i=1,2,3$ satisfy the following   
commutation relations:
$[A_i, A_j] = i \epsilon_{ijk} A_k$, which is identical to that satisfied
by the spin operators $S_k$ (or $S'_k$) of the channel $\Phi_1$ of
\reff{ch3}. This again proves the equivalence of the channels $\Phi_1$
and ${\widetilde{\Phi}}_3$.

As mentioned in the Introduction, 
the Holevo capacity and the minimum output entropy of the channel 
${\widetilde{\Phi}}_d$ have
been proved to be additive for any arbitrary dimension
$2\le d <\infty$ \cite{my, af, dhs, suff}. 
The equivalence of the channels $\Phi_1$
and ${\widetilde{\Phi}}_3$ implies
the validity of the additivity relations for the channel $\Phi_1$
as well.
Moreover, in \cite{suff} it was shown that the sufficient 
condition \reff{suffcond} was satisfied
for the channel ${\widetilde{\Phi}}_d$. Hence it is also satisfied
for the channel $\Phi_1$.

\section{Acknowledgements} 
We would like to thank A.S. Holevo and Y.M. Suhov for helpful
discussions. We are also grateful to A. Ekert for pointing out the 
connection between the channel $\Phi_{1/2}$ and the U-NOT gate.


\end{document}